\documentclass[12pt,a4paper]{article}
\usepackage{epsfig}
\begin{document}
\title{Probing signals of the littlest Higgs model via the $WW$ fusion
processes at the high energy $e^{+}e^{-}$ collider}
\author{Chong-Xing Yue, Wei Wang, Zheng-Jun Zong, and Feng Zhang\\
{\small  Department of Physics, Liaoning Normal University, Dalian
116029, China}\thanks{E-mail:cxyue@lnnu.edu.cn}\\}
\date{\today}

\maketitle
\begin{abstract}
In the framework of the littlest Higgs($LH$) model, we consider
the processes $e^{+}e^{-}\rightarrow\nu\bar{\nu}H^{0}$ and
$e^{+}e^{-}\rightarrow
W^{*}W^{*}\nu\bar{\nu}\rightarrow\nu\bar{\nu}t\bar{t}$, and
calculate the contributions of new particles to the cross sections
of these processes in the future high energy $e^{+}e^{-}$
collider($ILC$) with $\sqrt{S}=1TeV$. We find that, with
reasonable values of the free parameters, the deviations of the
cross sections for the processes
$e^{+}e^{-}\rightarrow\nu\bar{\nu}H^{0}$  from its $SM$ value
might be comparable to the future $ILC$ measurement precision. The
contributions of the light Higgs boson $H^0 $ to the process
$e^{+}e^{-}\rightarrow
W^{*}W^{*}\nu\bar{\nu}\rightarrow\nu\bar{\nu}t\bar{t}$ are
significant large in all of the parameter space preferred by the
electroweak precision data, which might be detected in the future
$ILC$ experiments. However, the contributions of the new gauge
bosons $B_{H}$ and $Z_{H}$ to this process are very small.

 \vspace{1cm}


\end{abstract}

\newpage
\noindent{\bf I. Introduction}

Little Higgs models[1,2,3] employ an extended set of global and
gauge symmetries in order to avoid the one-loop quadratic
divergences and thus provide a new method to solve the hierarchy
between the $TeV$ scale of possible new physics and the
electroweak scale $\nu=246GeV=(\sqrt{2}G_{F})^{-1/2}$. The key
feature of this type of models is that the Higgs boson is a
pseudo-Goldstone boson of a global symmetry which is spontaneously
broken at some higher scale $f$ and thus is naturally light.
Electroweak symmetry breaking($EWSB$) is induced by a
Coleman-Weinberg potential, which is generated by integrating out
the heavy degrees of freedom. This type of models can be regarded
as one of the important candidates of the new physics beyond the
standard model($SM$).

The next generation of high energy $e^{+}e^{-}$ linear
colliders($ILC'$s) are expected to operate at the center-of
-mass(c.m.) energy $\sqrt{S}=300GeV--1.5TeV$, which are required
to complement the probe of the new particles with detailed
measurement[4]. They will offer an excellent opportunity to study
the dynamics of the new physics with uniquely high precision. The
main production mechanism of the neutral Higgs boson in these
collider experiments are the Higgs-strahlung process
$e^{+}e^{-}\rightarrow ZH^{0}$ and the $WW$ fusion process
$e^{+}e^{-}\rightarrow W^* W^* \nu \overline{\nu}\rightarrow\nu
\overline{\nu}H^{0}$[5]. The cross section for the Higgs-strahlung
process scales as $1/S$ and dominates at low energies, while the
cross section for the $WW$ fusion process rises
$\log(S/m_{H}^{2})$ and dominates at high energies. It has been
shown that, for $\sqrt{S}\geq 500GeV$, the $WW$ fusion
contributions dominate the total cross section for the Higgs
production processes[6]. The $ZZ$ fusion process $e^{+}e^{-}
\rightarrow Z^* Z^* e^{+}e^{-}\rightarrow e^{+}e^{-}H^{0}$ can
also contribute to the Higgs boson production. However, the cross
section is suppressed by an order of magnitude compared to that
for the $WW$ fusion process, due to the ratio of the
$W^{\pm}e\nu_{e}$ coupling to the $Ze e $ coupling,
$4c_{W}^{2}=3$.

In Ref.[7], we have calculated the cross section of the
Higgs-strahlung process $e^{+}e^{-}\rightarrow ZH^{0}$ in the
context of the littlest Higgs($LH$) model[1]. We find that, in
most of the parameter space, the deviation of the total cross
section from its $SM$ value is larger than $5\%$, which may be
detected at the future $ILC$ experiment with $\sqrt{S}=500GeV$. In
this paper, we will consider the $WW$ fusion process
$e^{+}e^{-}\rightarrow W^* W^* \nu \overline{\nu}\rightarrow\nu
\overline{\nu}H^{0}$ and see whether the light Higgs boson
predicted by the $LH$ model can be detected via this process at
the future $ILC$ experiment with $\sqrt{S}=1TeV$.

It is well known that vector boson scattering processes can be
used to probe kinds of $EWSB$ mechanism at $TeV$ energies[8]. The
$WW$ fusion process $W^{+}W^{-}\rightarrow t\overline{t}$ could be
used to probe how the Higgs sector couples to fermions. Although
$QCD$ backgrounds make this process very difficult to observe at
the hadron colliders, it has been shown[9] that the signals of the
$SM$ Higgs sector could be established with good statistical
significance at the $ILC$ with $\sqrt{S}=1.5TeV$. In this paper,
we will study the $WW$ fusion process $W^{+}W^{-}\rightarrow
t\overline{t}$ at the future $ILC$ with $\sqrt{S}=1TeV$. In the
context of the $LH$ model, we calculate the contributions of the
light Higgs boson $H^{0}$ to this process and further calculate
the cross section for the process $e^{+}e^{-}\rightarrow W^* W^*
\nu \overline {\nu}\rightarrow \nu\overline{\nu}t\overline{t}$
using the effective $W$-boson approximation (EWA)[10]. We find
that the cross section of this process is very sensitive to the
free parameters of the $LH$ model and the possible signals of the
little Higgs boson $H^{0}$ should be detected at the future $ILC$
experiments with $\sqrt{S}=1TeV$.

The $LH$ model predicts the existence of the heavy gauge bosons,
such as $Z_{H}$ and $B_{H}$. We further study the contributions of
these new gauge bosons to the $WW$ fusion process
$e^{+}e^{-}\rightarrow W^* W^* \nu \overline {\nu}\rightarrow
\nu\overline{\nu}t\overline{t}$ in this paper. We find that the
contributions of gauge boson $Z_{H}$ exchange and $B_{H}$ exchange
to this process are very small in all of the parameter space
preferred by the electroweak precision data, which can not be
detected in the future $ILC$ experiments.

In the next section, we give the  couplings and masses of the new
particles predicted by the $LH$ model, which are related to our
calculation. In Sec.III we calculate the single production
cross-section of the light Higgs boson $H^{0}$ via the $WW$ fusion
process and compare our numerical result with that given in the
$SM$. The contributions of the little Higgs boson $H^{0}$ to the
process $W^{+}W^{-}\rightarrow t\overline{t}$ are studied in
Sec.IV. Using the EWA method, we further calculate the cross
section for the process $e^{+}e^{-}\rightarrow W^* W^* \nu
\overline{\nu}\rightarrow\nu \overline{\nu}t\overline{t}$
generated by $H^{0}$ exchange in this section. The possible
contributions of the heavy gauge bosons to the process
$e^{+}e^{-}\rightarrow W^* W^* \nu \overline{\nu}\rightarrow\nu
\overline{\nu}t\overline{t}$ are studied in Sec.V. Our conclusions
and discussions are given in Sec.VI.

\noindent{\bf II. The relevant coupling forms}

The $LH$ model[1] is one of the simplest and phenomenologically
viable models, which realizes the little Higgs idea. It consists
of a non-linear $\sigma$ model with a global $SU(5)$ symmetry and
a locally gauged symmetry $SU(2)_{1}\times U(1)_{1}\times
SU(2)_{2}\times U(1)_{2}$. The global $SU(5)$ symmetry is broken
down to its subgroup $SO(5)$ by a vacuum condensate $f\sim\Lambda
s/4\pi\sim TeV$, which results in fourteen massless Goldstone
bosons. Four of these particles are eaten by the $SM$ gauge
bosons, so that the locally gauged symmetry $SU(2)_{1}\times
U(1)_{1}\times SU(2)_{2}\times U(1)_{2}$ is broken down to its
diagonal subgroup $SU(2)\times U(1)$, identified as the $SM$
electroweak gauge group. The remaining ten Goldstone bosons
transform under the $SM$ gauge group as a doublet H and a triplet
$\Phi$. The doublet H becomes the $SM$ Higgs doublet, while the
triplet $\Phi$ is an addition to the $SM$ particle contents. This
breaking scenario also gives rise to the new gauge bosons
$W_{H}^{\pm}, B_{H}, Z_{H}$.

In the $LH$ model, the light Higgs boson acquires the mass squared
parameter at two-loop as well as at one-loop from the
Coleman-Weinberg potential. Its mass is protected from the
one-loop quadratic divergence by a few new particles with the same
statistics as the corresponding $SM$ particles. The new heavy
gauge bosons $W_{H}^{\pm}, B_{H}, Z_{H}$ cancel the one-loop
quadratic divergence generated by the $SM$ gauge boson $W$ and $Z$
loops. New heavy scalar $\Phi$  cancels that generated by the
Higgs self-interaction. A new vector-like top quark $T$ is also
needed to cancel the divergences from the top quark Yukawa
interactions. Furthermore, these new particles might produce
characteristic signatures at the present and future collider
experiments[7,11,12,13]. Certainly, these new particles can
generate significant corrections to some observables and thus the
precision measurement data can give severe constraints on this
kind of models[11,14,15,16].

In the $LH$ model, the coupling expressions of the Higgs boson
$H^{0}$, which are related to our calculation, can be written
as[11]:
\begin{eqnarray}
&&g^{H^{0}W^{+}_{\mu}W^{-}_{\nu}}=\frac{ie^{2}\nu
g_{\mu\nu}}{2s_{W}^{2}}
[1-\frac{\nu^{2}}{3f^{2}}+\frac{1}{2}(c^{2}-s^{2})^{2}\frac{\nu^{2}}{f^{2}}
-12\frac{\nu'}{\nu}],\\
&&g^{H^{0}W^{+}_{H\mu}W^{-}_{H\nu}}=-\frac{ie^{2}\nu
}{2s_{W}^{2}}g_{\mu\nu},\ \ \ \ \ \ \
g^{H^{0}W^{+}_{\mu}W^{-}_{H\nu}}=-\frac{ie^{2}\nu
g_{\mu\nu}}{2s_{W}^{2}}\frac{(c^{2}-s^{2})}{2sc},\\
&&g^{H^{0}t\overline{t}}=-\frac{im_{t}}{\nu}
[1-4(\frac{\nu'}{\nu})^{2}+2\frac{\nu'}{f}-\frac{2}{3}(\frac{\nu}{f})^{2}+
\frac{\nu^{2}}{f^{2}}x_{L}(1+x_{L})].
\end{eqnarray}
Where $s_{W}=\sin\theta_{W}$, $\theta_{W}$ is the Weinberg angle,
$\nu'$ is vacuum expectation value(VEV) of the triplet scalar $
\Phi$. $c(s=\sqrt{1-c^{2}})$ is the mixing parameter between
$SU(2)_{1}$ and $SU(2)_{2}$ gauge bosons and the mixing parameter
$c'(s'=\sqrt{1-c^{'2}})$ comes from the mixing  between $U(1)_{1}$
and $U(1)_{2}$ gauge bosons. Using these mixing parameters, we can
represent the $SM$ gauge coupling constants as $g=g_{1}s=g_{2}c$
and $g'=g'_{1}s'=g'_{2}c'$. The mixing parameter between the $SM$
top quark t and the vector-like top quark T is defined as
$x_{L}=\lambda_{1}^{2}/(\lambda_{1}^{2}+\lambda_{2}^{2})$ , in
which $\lambda_{1}$ and $\lambda_{2}$ are the Yukawa coupling
parameters.

Taking account of the gauge invariance of the Yukawa couplings and
the $U(1)$ anomaly cancellation, the relevant couplings of the
gauge bosons $W$, $W_{H}^{\pm}$, $B_{H}$, and $ Z_{H}$ to ordinary
particles can be written as in the $LH$ model:
\begin{eqnarray}
&&g_{L}^{W\nu e}=\frac{ie}{\sqrt{2}s_{W}}
[1-\frac{\nu^{2}}{2f^{2}}c^{2}(c^{2}-s^{2})],\hspace{0.5cm}g_{R}^{W\nu e}=0;\\
&&g_{L}^{W_{H}\nu e}=-\frac{ie}{\sqrt{2}s_{W}} \frac{c}{s}, \hspace{0.5cm}
g_{R}^{W_{H}\nu e}=0;\\
&&g_{L}^{Wtb}=\frac{ie}{\sqrt{2}s_{W}}[1-\frac{\nu^{2}}{2f^{2}}
(x_{L}^{2}+c^{2}(c^{2}-s^{2}))],\hspace{0.5cm}
g_{R}^{Wtb}=0;\\
&&g^{B_{H}W^{+}_{\mu}W_{\nu}^{-}}=-\frac{ec_{W}}{s^{2}_{W}}
[\frac{\nu^{2}}{f^{2}}\frac{5}{2}s'c'(c'^{2}-s'^{2})],
\hspace{0.5cm}g^{Z_{H}W^{+}_{\mu}W_{\nu}^{-}}=\frac{e}{2s_{W}}
[\frac{\nu^{2}}{f^{2}}sc(c^{2}-s^{2})];\\
&&g_{L}^{B_{H}t\bar{t}}=\frac{e}{6c_{W}s'c'}(\frac{2}{5}-c'^{2}),
\hspace{0.5cm}g_{R}^{B_{H}t\bar{t}}
=\frac{2e}{3c_{W}s'c'}[(\frac{2}{5}-c'^{2})-\frac{3}{20}x_{L}];\\
&&g_{L}^{Z_{H}tt}=\frac{ec}{2s_{W}s},\hspace{1cm}
g_{R}^{Z_{H}tt}=0.
\end{eqnarray}

To obtain our numerical results, we write the masses of the
relevant particles as:
\begin{eqnarray}
M_{W}^{2}&=&m_{W}^{2}[1-\frac{\nu^{2}}{f^{2}}(\frac{1}{6}+\frac{1}{4}
(c^{2}-s^{2})^{2})+ 4(\frac{\nu'^{2}}{\nu^{2}})],\\
M_{W_{H}}^{2}&\approx&
m_{W}^{2}(\frac{f^{2}}{s^{2}c^{2}\nu^{2}}-1),\hspace{0.5cm}
M_{B_{H}}^{2}\approx
\frac{m_{W}^{2}s^{2}_{W}}{c_{W}^{2}}[\frac{f^{2}}{5s^{'2}c^{'2}\nu^{2}}-1],\\
M_{Z_{H}}^{2}&\approx&M_{W_{H}}^{2},
\end{eqnarray}
where $m_{W}=g\nu/2$ is the mass of the $SM$ gauge boson $W$. From
above equations, we can see that, at the order of $\nu^{2}/f^{2}$,
the $B_{H}$ mass $M_{B_{H}}$ and the $Z_{H}$ mass $M_{Z_{H}}$
mainly depend on the free parameters ($f$, $c'$) and ($f$, $c$),
respectively. In general, the heavy photon $B_{H}$ is
substantially lighter than the gauge boson $Z_{H}$. Considering
the constraints of the electroweak precision data on the free
parameters $f,\  c,$ and $c'$, the value of the ratio $
M_{B_{H}}^{2}/M_{Z_{H}}^{2}$ can be further reduced.

In the following calculation, we will take the mass of the light
Higgs boson $m_{H}=115GeV$. In this case, the possible decay modes
of $H^{0}$ are $b\overline{b}$, $c\overline{c}$, $l\overline{l}$
[$l=\tau, \mu$ or $e$], $gg$ and $\gamma\gamma$. However, the
total decay width $\Gamma_{H}$ is dominated by the decay channel
$H^{0}\rightarrow b\overline{b}$. In the $LH$ model, $\Gamma_{H}$
is modified from that in the $SM$ by the order of $\nu^{2}/f^{2}$
and has been studied in Ref.[13].

Considering the electroweak precision data constraints, the
$B_{H}$ mass $M_{B_{H}}$ is not too heavy and can be  allowed to
be in the range of a few hundred $GeV$[14]. For the decay channels
$B_{H}\rightarrow\overline{t}t$ and $B_{H}\rightarrow ZH$, we can
not neglect the final state masses.  The electroweak precision
data constrain the $Z_{H}$ mass $M_{Z_{H}}$ to be no smaller than
about $1TeV$. Thus, for all of the $Z_{H}$ decay channels, we can
neglect the final state masses. The total decay widths
$\Gamma_{Z_{H}}$ and $\Gamma_{B_{H}}$ of the gauge bosons $Z_{H}$
and $B_{H}$ have been discussed in Refs.[13,14]. It is easily to
know that $\Gamma_{B_{H}}$ is sensitive to the free parameters $f$
and $c'$, while $\Gamma_{Z_{H}}$ is sensitive to the free
parameters $f$ and $c$.

Global fits to the electroweak precision data produce rather
severe constraints on the parameter space of the $LH$ model[14,
15]. However, if the $SM$ fermions are charged under
$U(1)_{1}\times U(1)_{2}$, the constraints become relaxed. The
scale parameter $f=1\sim 2TeV$ is allowed for the mixing
parameters $c$, $c'$, and $x_{L}$ in the ranges of $0\sim 0.5$,
$0.62\sim 0.73$, and $0.3\sim0.6$, respectively[16]. Taking into
account the constraints on the free parameters $f$, $c$, $c'$ and
$x_{L}$, we will give our numerical results in the following
sections.

\noindent{\bf III. The $WW$ fusion process
$e^{+}e^{-}\rightarrow\nu \overline{\nu}H^{0}$ in the $LH$ model }

A future $ILC$ will measure the production cross section of a
light Higgs boson via $WW$ fusion with percent-level precision[4].
Furthermore, in the $ILC$ experiments with $\sqrt{S}\geq 500GeV$,
the $WW$ fusion process $e^{+}e^{-}\rightarrow\nu\overline{\nu}
H^{0}$ dominates single production of the Higgs boson $H^{0}$[6].
Thus, it is very interesting to study this process in the popular
specific models beyond the $SM$.

In the $SM$, the production cross section for the process
$e^{+}e^{-}\rightarrow\nu \overline{\nu}H^{0}$ can be generally
written as[5]:
\begin{equation}
\sigma^{SM}=\frac{G_{F}^{3}m_{W}^{4}}{4\sqrt{2}\pi^{3}}\int^{1}_{x_{H}}dx
\int^{1}_{x}\frac{dy F(x,y)}{[1+(y-x)/x_{W}]^{2}}
\end{equation}
with
\begin{equation}
F(x,y)=(\frac{2x}{y}-\frac{1+3x}{y^{2}}+\frac{2+x}{y}-1)[\frac{z}{1+z}
-\ln(1+z)]+\frac{xz^{2}(1-y)}{y^{3}(1+z)},
\end{equation}
where $x_{H}=m_{H}^{2}/S$, $ x_{W}=m_{W}^{2}/S$, and
$z=y(x-x_{H})/(xx_{W})$.

Compared with the $WW$ fusion process $e^{+}e^{-}\rightarrow\nu
\overline{\nu}H^{0}$ in the $SM$, this process in the $LH$ model
receives additional contributions from the heavy gauge bosons
$W_{H}^{\pm}$, proceed through the Feynman diagrams depicted in
Fig.1. Furthermore, the modification of the relations among the
$SM$ parameters and the precision electroweak input parameters,
and the correction terms to the $SM$ $We\nu_{e}$ coupling can also
produce corrections to this process.

\begin{figure}[htb]
\vspace{-8cm}
\begin{center}
\epsfig{file=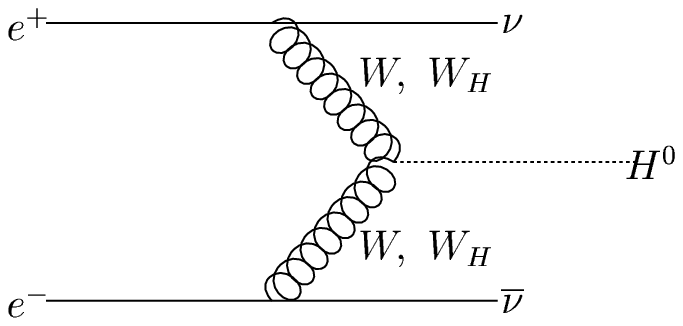,width=520pt,height=700pt} \vspace{-13.5cm}
\hspace{0.5cm} \caption{Feynman diagrams for the $ WW $
 fusion process $e^{+}e^{-}\rightarrow\nu \overline{\nu}H^{0}$
  in the $LH$ model.} \label{ee}
\end{center}
\end{figure}

In the $LH$ model, the relation among the Fermi coupling constant
$G_{F}$, the  gauge boson $W$ mass $m_{W}$ and the fine structure
constant $\alpha$ can be written as[16]:
\begin{equation}
\frac{G_{F}}{\sqrt{2}}=\frac{\pi\alpha}{2m_{W}^{2}s_{W}^{2}}[1-c^{2}
(c^{2}-s^{2})\frac{\nu^{2}}{f^{2}}
+2c^{4}\frac{\nu^{2}}{f^{2}}-\frac{5}{4}(c^{'2}-s^{'2})\frac{\nu^{2}}
{f^{2}}].
\end{equation}
So we have
\begin{equation}
\frac{e^{2}}{s_{W}^{2}}=\frac{4\sqrt{2}G_{F}m_{W}^{2}}{[1-c^{2}
(c^{2}-s^{2})\frac{\nu^{2}}{f^{2}}
+2c^{4}\frac{\nu^{2}}{f^{2}}-\frac{5}{4}(c^{'2}-s^{'2})\frac{\nu^{2}}{f^{2}}]}.
\end{equation}
In the following numerical estimation, we will take $G_{F}=1.16637
\times 10^{-5}GeV^{-2},\ m_{Z}=91.18GeV$ and $m_{W}=80.45GeV$[17]
as input parameters and use them to represent the other $SM$
parameters. The $c.m.$ energy $\sqrt{S}$ of the future $ILC$
experiments is assumed as $\sqrt{S}=1TeV$.

\begin{figure}[htb]
\vspace*{-0.5cm}
\begin{center}
\epsfig{file=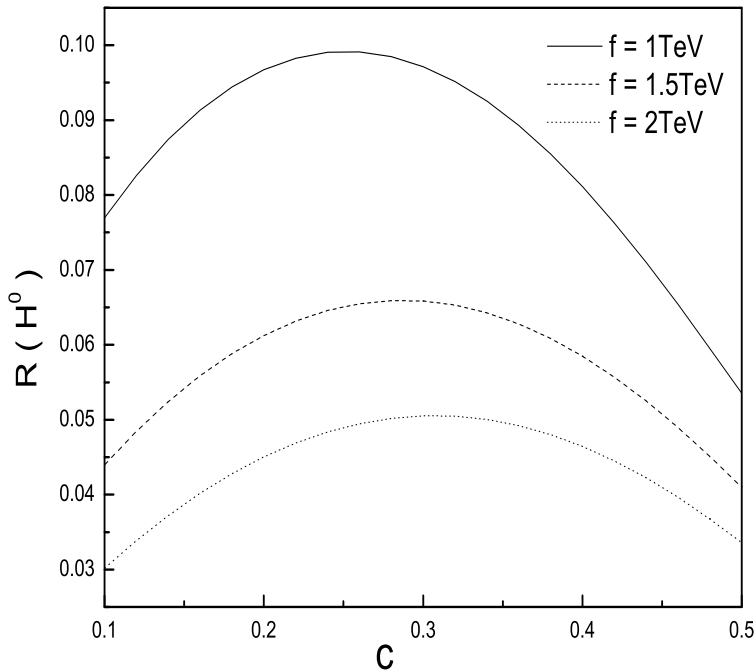,width=320pt,height=300pt} \vspace{-1cm}
\hspace{0.5cm} \caption{The parameter $R(H^{0})$ as a function of
the mixing parameter $c$ for $\nu'/\nu=\nu/5f$ \hspace*{1.8cm}and
three values of the scale parameter $f$.} \label{ee}
\end{center}
\end{figure}

Except for the $SM$ input parameters, there are three free
parameters in the expression of the relative correction parameter
$R(H^{0})=\Delta\sigma/\sigma^{SM}$ with
$\Delta\sigma=\sigma^{LH}-\sigma^{SM}$: the mixing parameter $c$,
the scale parameter $f$, and the triplet scalar VEV $\nu'$. In
order to obtain the correct $EWSB$ vacuum and avoid giving a
$TeV$-scale VEV to the scalar triplet $\Phi$, we should have that
the value of $\nu'/\nu$ is smaller than $\nu/4f$[1, 11]. In Fig.2,
we plot the relative correction parameter $R(H^{0})$ as a function
of the mixing parameter $c$ for $\nu'/\nu=\nu/5f$ and three values
of the scale parameter $f$. From Fig.2, we can see that the value
of $R(H^{0})$ decreases as $f$ increasing, which is consistent
with the conclusions for the corrections of the $LH$ model to
other observables. If we assume $f=1TeV$, the value of the
relative correction parameter $R(H^{0})$ is larger than $5.4\%$ in
all of the parameter space preferred by the electroweak precision
data. For $f\geq 2TeV$, $R(H^{0})$ is smaller than $5\%$ in most
of the parameter space of the $LH$ model.

To see the effects of the varying triplet scalar VEV $\nu'$ on the
relative correction parameter $R(H^{0})$, we take $f=1TeV$, which
means $\nu'/\nu\leq \nu/4f=0.061$, and plot $R(H^{0})$ as a
function of $\nu'/\nu$ in Fig.3 for three values of the mixing
parameter $c$. One can see from Fig.3 that $R(H^{0})$ is not
sensitive to the ratio $\nu'/\nu$, compared with the mixing
parameter $c$. For $f=1TeV$ and $\nu'/\nu\leq 0.06$, the value of
$R(H^{0})$ is larger than $4\%$ and $6\%$ for the mixing parameter
$c=0.1$ and $0.3$, respectively.

\begin{figure}[htb] \vspace*{0cm}
\begin{center}
\epsfig{file=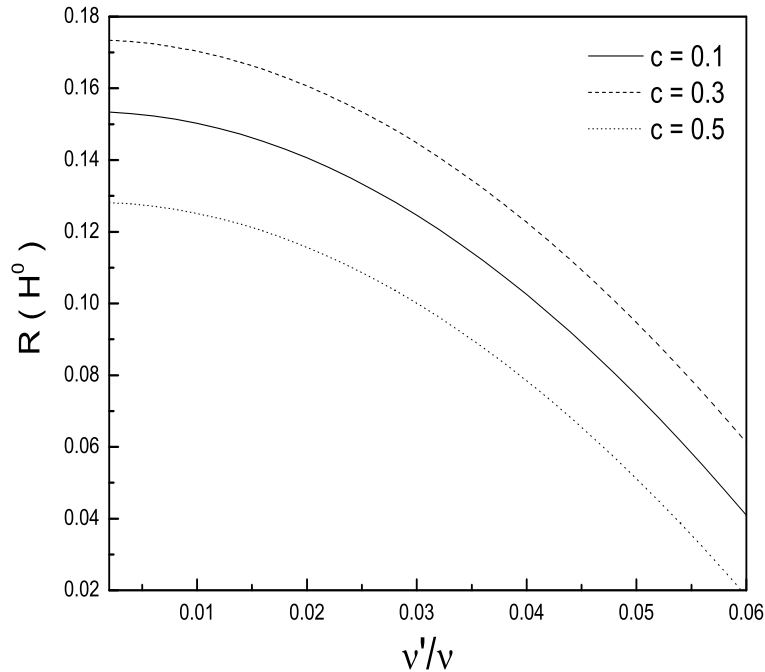,width=320pt,height=300pt} \vspace{-1cm}
\hspace{0.5cm} \caption{The parameter $R(H^{0})$ as a function of
$\nu'/\nu$ for $f=1TeV$ and three values of \hspace*{1.8cm}the
mixing parameter $c$.} \label{ee}
\end{center}
\end{figure}

In general, the $LH$ model can produce corrections to single
production of the light Higgs boson $H^{0}$ via the $WW$ process
$e^{+}e^{-}\rightarrow \nu\bar{\nu}H^{0}$ at the future $ILC$
experiments. Our results show that the correction effects on the
production cross section can be significant large in all of the
parameter space of the $LH$ model. Even if we take account of the
constraints of the electroweak precision data on the free
parameters of the $LH$ model, the value of the relative correction
parameter $R(H^{0})$is generally larger than $5\%$. A future $ILC$
will measure the production cross section of a light Higgs boson
from Higgs-strahlung or $WW$ fusion process with percent-level
precision, as well as the important branching fractions with
few-percent precision[4,18]. Thus, correction effects of the $LH$
model on the $WW$ fusion process $e^{+}e^{-}\rightarrow
\nu\bar{\nu}H^{0}$ might be comparable to the future $ILC$
measurement precision.

\noindent{\bf IV. The Higgs boson $H^{0}$ and the process
$W^{+}W^{-}\rightarrow t\overline{t}$ in the $LH$ model}

The production cross section of the process $W^{+}W^{-}
\rightarrow t\overline{t}$ generated by the Higgs boson $H^{0}$ is
sensitive to the terms proportional to the coupling parameters
$(g^{H^{0}f\bar{f}})^{2}$ and $g^{H^{0}f\bar{f}}$ of the Higgs
boson $H^{0}$ to fermions, which come from pure Higgs
contributions and the interference with non-Higgs contributions,
respectively. Thus, the process $W^{+}W^{-}\rightarrow
t\overline{t}$ could be used to probe how the Higgs sector couples
to fermions. Although $QCD$ backgrounds make this process very
difficult to observe at the hadron colliders, the signals of the
Higgs sector could be established with good statistical
significance at the high energy $ILC$ experiments[8,9,19]. In this
section, we consider the contributions of the Higgs boson $H^{0}$
to this process in the context of the $LH$ model and calculate the
relative deviations from the $SM$ prediction.

The subprocess $W^{+}W^{-}\rightarrow t\overline{t}$ can be
effectively realized via gauge boson $W$ radiation from initial
fermion lines:
\begin{equation}
e^{+}e^{-}\rightarrow \nu\overline{\nu}W^{*}W^{*}\rightarrow
\nu\overline{\nu}t\overline{t},
\end{equation}
which was first calculated in Ref.[20] by using the EWA
method[10]. In the approach of an effective Lagrangian, Ref.[21]
has extensively studied this process. Effects of the models of
strong interaction $EWSB$ on the subprocess $W^{+}W^{-}\rightarrow
t\overline{t}$ were discussed in Ref.[8].

For large $\sqrt{\hat{S}}$, which is the c.m. energy of the
subprocess $W^{+}W^{-}\rightarrow t\overline{t}$ in the $ILC$ with
the c.m. energy $\sqrt{S}=1 TeV$, the longitudinal polarization
vector of gauge bosons $W^{\pm}$ can be approximately expressed by
$\varepsilon_{0}^{\mu}(k)\approx k^{\mu}/m_{W}
+O(m_{W}/\sqrt{\hat{S}})$. The term $k^{\mu}/m_{W}$ produces the
leading contributions to the cross section $\hat{\sigma}(\hat{S})$
for the subprocess $W^{+}W^{-}\rightarrow t\overline{t}$, which
are proportional to $(m_{t}/m_{W})^{4}$, while the sub-leading
contributions generated by the term $O(m_{W}/\sqrt{\hat{S}})$ are
suppressed by a factor $m_{t}^{2}/\hat{S}$. Thus, the production
cross section $\hat{\sigma}(\hat{S})$ for the subprocess
$W^{+}W^{-}\rightarrow t\overline{t}$ is well approximated by
taking only the longitudinal polarized W's at the parton level
reaction and assuming $\hat{S}\geq m_{W}^{2}$[10,20,21,22].
However, in this paper, we want to calculate the contributions of
the Higgs boson $H^{0}$ to the cross section for the subprocess
$W^{+}W^{-}\rightarrow t\overline{t}$ in the $LH$ model and
compare our numerical result with that in the $SM$. Thus, we will
include all polarizations for the gauge bosons $W^{\pm}$ in our
calculation of the production cross section
$\hat{\sigma}(\hat{S})$.

In the $LH$ model, the production cross section
$\hat{\sigma}(\hat{S})$ for the subprocess
$W^{+}_{\lambda_{+}}W^{-}_{\lambda_{-}}\rightarrow t\overline{t}$
generated by the Higgs boson $H^{0}$ can be written as:
\begin{eqnarray}
\hat{\sigma}(\hat{S})\nonumber
&=&\frac{3\pi\alpha^{2}A^{2}}{2s_{W}^{4}}\cdot
m_{t}^{2}X_{H}^{2}\beta_{t}^{3}\mid\varepsilon^{W^{+}}_{\lambda_{+}}\cdot
\varepsilon^{W^{-}}_{\lambda_{-}}\mid^{2}\\\nonumber
&&+\frac{3\pi\alpha^{2}AB^{2}}{8s_{W}^{4}}\cdot
\frac{m_{t}^{2}}{\hat{S}}X_{H}\beta_{t}(1-\beta_{t}^{2})\\\nonumber
&&\cdot\{\mid\varepsilon^{W^{+}}_{\pm}\cdot
\varepsilon^{W^{-}}_{\pm}\mid^{2}[-1+\frac{1+\beta_{t}^{2}}{2\beta_{t}^{2}}L]\\&&+4
\mid\varepsilon^{W^{+}}_{0}\cdot
\varepsilon^{W^{-}}_{0}\mid^{2}\cdot[-\frac{1+\beta_{t}^{2}}{1-\beta_{t}^{2}}+
\frac{(1-\beta_{t}^{2})}{2\beta_{t}^{2}}L]\}
\end{eqnarray}

with
\begin{equation}
A=[1-\frac{\nu^{2}}{3f^{2}}+\frac{\nu^{2}}{2f^{2}}(c^{2}-s^{2})-12(\frac{\nu^{'}}
{\nu})^{2}]\cdot
[1-4(\frac{\nu^{'}}{\nu})^{2}+2\frac{\nu'}{f}-\frac{2}{3}\frac{\nu^{2}}{f^{2}}
+\frac{\nu^{2}}{f^{2}}
x_{L}(1+x_{L})],
\end{equation}
\begin{equation}
B=1-\frac{\nu^{2}}{2f^{2}}[c^{2}(c^{2}-s^{2})+x_{L}^{2}],\hspace{1.0cm}
L= \ln(\frac{1+\beta_{t}}{1-\beta_{t}}),
\end{equation}
\begin{equation}
\beta_{t}=\sqrt{1-\frac{4m_{t}^{2}}{\hat{S}}},\hspace{1.0cm}
X_{H}=\frac{\hat{S}-m_{H}^{2}}{(\hat{S}-m_{H}^{2})^{2}+m_{H}^{2}\Gamma_{H}^{2}}.
\end{equation}
The second term of $Eq.(18)$ comes from the interference effects
of the s-channel $H^{0}$ exchange with the t-channel $b$ quark
exchange. Due to the orthogonality properties of the polarizations
vectors $\varepsilon^{W^{\pm}}_{\lambda_{\pm}}$ of the gauge
bosons $W^{\pm}$, there is no interference between the transverse
and the longitudinal polarizations. So we have
\begin{equation}
\mid\varepsilon^{W^{+}}_{\pm}\cdot
\varepsilon^{W^{-}}_{\mp}\mid^{2}=0,\
\mid\varepsilon^{W^{+}}_{\pm}\cdot
\varepsilon^{W^{-}}_{\pm}\mid^{2}=1,\\\\\\
\mid\varepsilon^{W^{+}}_{0}\cdot \varepsilon^{W^{-}}_{0}\mid^{2}=
\frac{(1+\beta_{W}^{2})^{2}}{(1-\beta_{W}^{2})^{2}}
\end{equation}
 with $ \beta_{W}=\sqrt{1-4m^{2}_{W}/\hat{S}}$.

In general, the cross section $\sigma(S)$ for the process
$e^{+}e^{-} \rightarrow \nu\overline {\nu} W^{*} W^{*} \rightarrow
\nu \overline{\nu} t\overline{t}$ can be obtained by folding the
cross section $\hat{\sigma}(\hat{S})$ for the subprocess
$W^{+}_{\lambda_{+}}W^{-}_{\lambda_{-}}\rightarrow t \overline{t}$
with the $W^{\pm}$ distribution functions
$f^{W^{\pm}}_{\lambda_{\pm}}$ with helicities $\lambda_{\pm}$:
\begin{equation}
\sigma(S)=\Sigma_{\lambda_{+},\lambda_{-}}\int^{1}_{2m_{t}/\sqrt{s}}2xdx\int^{1}_{x^{2}}
\frac{dx_{+}}{x_{+}}f^{W^{+}}_{\lambda_{+}}(x_{+})f^{W^{-}}_{\lambda_{-}}
(\frac{x^{2}}{x_{+}})\hat{\sigma}(\hat{S}),
\end{equation}
where $x^{2}=\hat{S}/S$, the helicities $\lambda_{\pm}$ of the
gauge bosons $W^{\pm}$ each run over 1, 0, -1. In our
calculations, we use the full distributions given by Refs.[10,20]
for $f^{W^{\pm}}_{\lambda_{\pm}}(x)$ and include all polarizations
for the gauge bosons $W^{\pm}.$

To discuss the deviation of the production cross section
$\sigma_{1}^{LH}(t\overline{t})$ for the process
$e^{+}e^{-}\rightarrow \nu\overline{\nu}H^{0}\rightarrow
\nu\overline{\nu} t\overline{t}$ in the $LH$ model from its $SM$
value, we define the relative correction parameter:
$R_{1}(t\overline{t})=\Delta\sigma_{1}(t\overline{t})/\sigma_{1}^{SM}(t\overline{t})$
with
$\Delta\sigma_{1}(t\overline{t})=\sigma_{1}^{LH}(t\overline{t})-\sigma_{1}^{SM}
(t\overline{t})$, in which $\sigma_{1}^{SM} (t\overline{t})$
denotes the production cross section for this process
 in the $SM$. Obviously, the value of the relative
correction parameter $R_{1}(t\bar{t})$ increases as the scale
parameter $f$ decreasing. Considering the constraints from the
precision measurement data on the free parameters of the $LH$
model, we will assume $f\geq 1TeV$ in the following numerical
estimation.

\begin{figure}[htb]
\begin{minipage}[t]{0.5\linewidth}
\centering
\includegraphics[width=8cm]{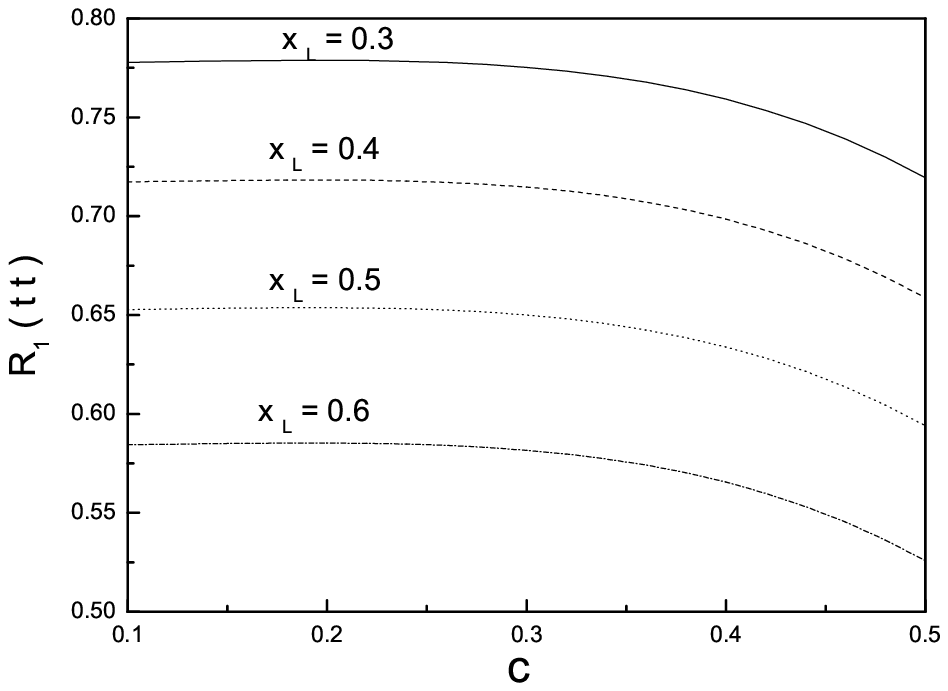}
\caption{\footnotesize The parameter $R_{1}(t\bar{t})$ as function
of $c$ \hspace*{1.9cm}for $\nu'/\nu=\nu/5f$, $f=1TeV$ and four
\hspace*{1.9cm}values of the mixing parameter
$x_{L}$.}\label{fig:side:a}
\end{minipage}%
\hspace{0.5cm}
\begin{minipage}[t]{0.5\linewidth}
\centering
\includegraphics[width=8cm]{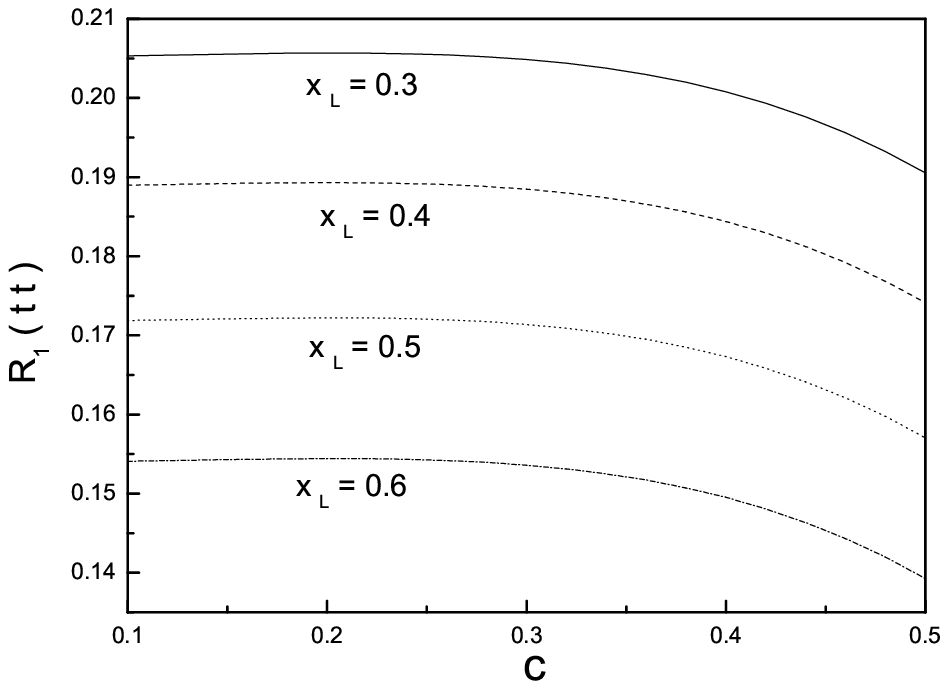}
\caption{\footnotesize Same as Fig.4 but for
$f=2TeV$.}\label{fig:side:b}
\end{minipage}%
\end{figure}

In Fig.4(Fig.5), we plot the relative correction parameter
$R_{1}(t\bar{t})$ as a function of the mixing parameter $c$ for
$\nu'/\nu=\nu/5f,\ f=1TeV(2TeV)$, and four values of the mixing
parameter $x_{L}$. From these figures, we can see that the
relative correction parameter $R_{1}(t\bar{t})$ increases as the
mixing parameter $x_{L}$ decreasing and is insensitive to the
mixing parameter $c$. For $f=1TeV$, the value of $R_{1}(t\bar{t})$
is larger than $50\%$ in all of the parameter space preferred by
the electroweak precision data. Even if we assume the scale
parameter $f=2TeV$, the value of $R_{1}(t\bar{t})$ is larger than
$10\%$.

The relative correction parameter $R_{1}(t\bar{t})$ is plotted in
Fig.6(Fig.7) as a function of  $\nu'/\nu$ for $f=1TeV(2TeV)$,
$x_{L}=0.5$ and three values of the mixing parameter $c$. From
Fig.6 and Fig.7, one can see that the value of $R_{1}(t\bar{t})$
increases as $\nu'/\nu$ increasing. As long as the scale parameter
 $f\leq 2TeV$, its value is larger than $10\%$ in all of the parameter
 space preferred by the electroweak precision data.

\begin{figure}[htb]
\begin{minipage}[t]{0.5\linewidth}
\centering
\includegraphics[width=8cm]{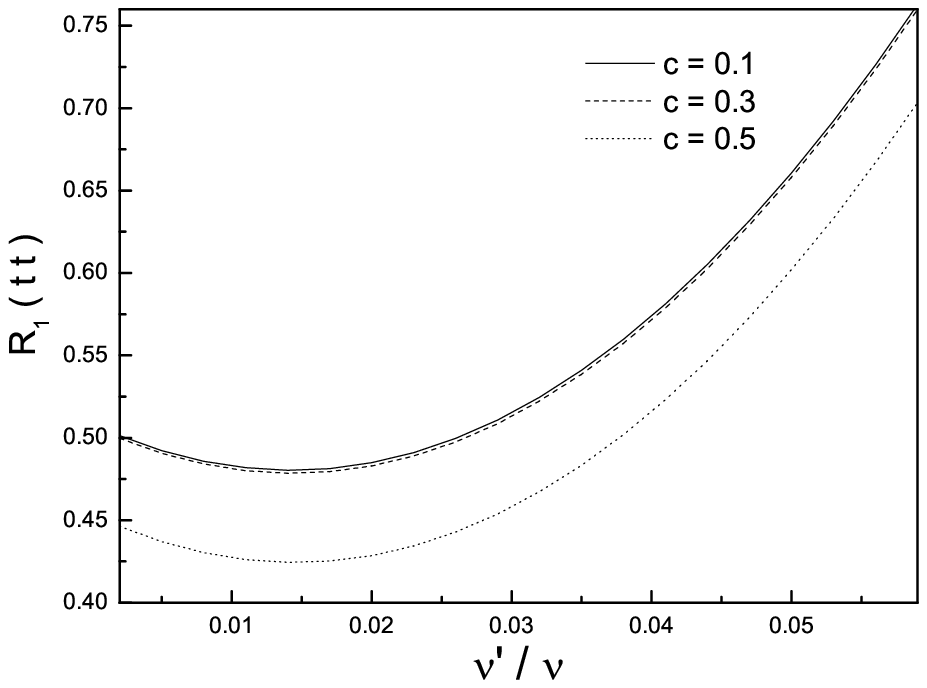}
\caption{\footnotesize The parameter $R_{1}(t\bar{t})$ as a
function of \hspace*{1.9cm}$\nu'/ \nu $ for $f=1TeV$, $x_{L}=0.5$
and three \hspace*{1.9cm}values of the mixing parameter
$c$.}\label{fig:side:a}
\end{minipage}%
\hspace{0.5cm}
\begin{minipage}[t]{0.5\linewidth}
\centering
\includegraphics[width=8cm]{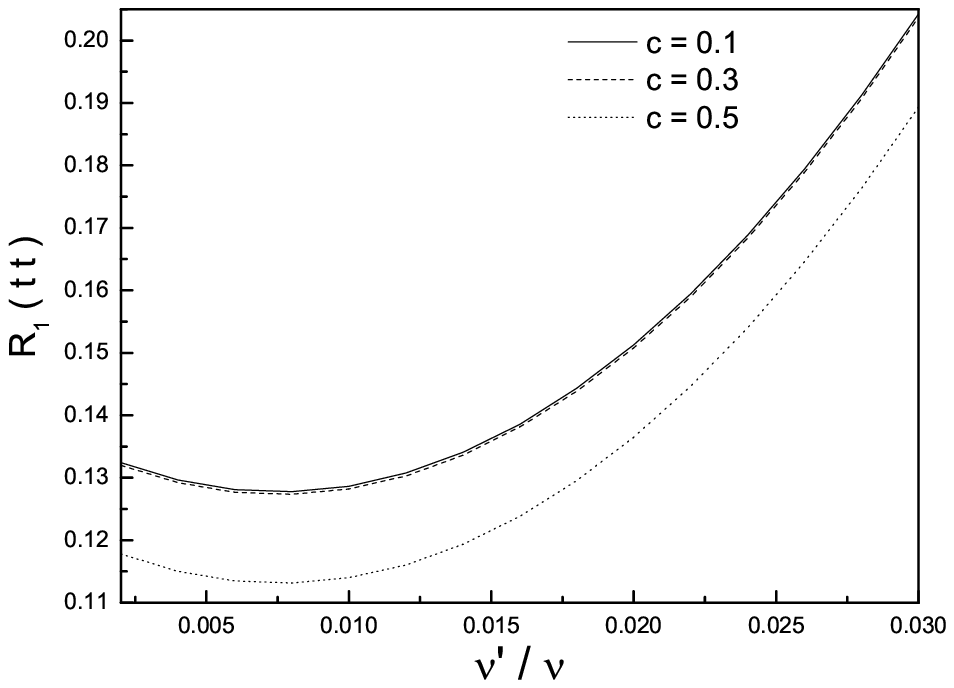}
\caption{\footnotesize Same as Fig.6 but for
$f=2TeV$.}\label{fig:side:b}
\end{minipage}%
\end{figure}

Using the EWA method[10], we have calculated the contributions of
the light Higgs boson to the process $e^{+}e^{-}\rightarrow
W^{*}W^{*}\nu\overline{\nu}\rightarrow
\nu\overline{\nu}t\overline{t}$. In our numerical estimation, we
have included all contributions of the longitudinal and transverse
$W$ boson components and taken the c.m. energy $\sqrt{S}=1TeV$.
However, the production cross section given by the EWA method is
larger than the exact result by a factor $2$ to $5$, which depends
on the considered c.m. energy and the Higgs boson mass[18].
Furthermore, Ref.[19] has shown that, at high energy $e^{+}e^{-}$
colliders with c.m. energies of $1.5TeV$ or above, the effective
$WW$ fusion calculation approximates well the exact result. Using
the computer code NextCalibur[23], we have check our numerical
results and find that, for $\sqrt{S}=1TeV$ and $m_{H}=115GeV$, the
values of the relative correction parameter $R_{1}(t\bar{t})$
shown in Fig.4-Fig.7 are approximately suppressed by a factor
$1/2.5$. Thus, we expect that, as long as $f\leq 1.5TeV$, the
value of $R_{1}(t\bar{t})$ is larger than $10\%$ in all of the
parameter space preferred by the electroweak precision data.

Due to the missing momenta in the longitudinal and transverse
directions, only the final $6-$jet events in which both the top
and antitop decay into a $b$ quark plus two additional quarks can
be fully reconstructed experimentally[19]. The signal of the
$t\bar{t}$ production via the process $e^{+}e^{-}\rightarrow
W^{*}W^{*}\nu\overline{\nu}\rightarrow \nu\overline{\nu}
t\overline{t}$ should be $6-$jet events associated with large
missing energy. The most dangerous backgrounds to this signal are
direct $t\bar{t}$ production via the process
$e^{+}e^{-}\rightarrow t\overline{t}$ and $e^{+}e^{-}
t\overline{t}$ production via the process $e^{+}e^{-}\rightarrow
e^{+}e^{-}t\overline{t}$. The former background can be efficiently
reduced by choosing the jet association that gives the best fit to
the reconstructed $t$ and $W$ masses and keeping events within
five standard deviations of the expected values, while the latter
background can be reduced by requiring the missing transverse
energy in the event to be greater than $50GeV$[19]. Thus, the
correction effects of the light Higgs boson on the production of
the top quark pair via the process $e^{+}e^{-}\rightarrow
W^{*}W^{*}\nu\overline{\nu}\rightarrow
\nu\overline{\nu}t\overline{t}$ should be detected at the future
$ILC$ experiments with $\sqrt{S}=1TeV$.

\noindent{\bf V. The heavy gauge bosons and the process
$e^{+}e^{-}\rightarrow \nu\overline{\nu}W^{*}W^{*}\rightarrow
\nu\overline{\nu}t\overline{t}$}

The process $e^{+}e^{-}\rightarrow \nu\overline{\nu}
W^{*}W^{*}\rightarrow \nu\overline{\nu}t\overline{t}$ is one of
the dominant production processes of the top quark pairs in the
future $ILC$ experiments. It is expected that there will be
thousands of $t\overline{t}$ pair events produced via $ WW $
fusion process at the future $ILC$ experiments with
$\sqrt{S}=1TeV$ and a yearly integrated luminosity
$\pounds=100fb^{-1}$. The $ILC$ will allow the couplings of the
longitudinal gauge bosons $W^{\pm}$ to the top quark to be very
accurately determined[20]. Thus, this process is very sensitive to
$EWSB$ mechanism and should be carefully studied within some
popular specific models beyond the $SM$.

In the $SM$, the subprocess $W^{+}W^{-}\rightarrow t\overline{t}$
can proceed through the t-channel b quark exchange and the
s-channel $\gamma,\ Z,\ H^{0}$ exchanges, which has been
extensively studied in Refs.[20,24]. In the $LH$ model, except for
the contributions of the light Higgs boson $ H^0$, this process
receives additional contributions from the heavy photon $B_{H}$
exchange and the new $SU(2)$ gauge boson $Z_{H}$ exchange in the
s-channel. In this section, we will consider the contributions of
the gauge bosons $B_{H}$ exchange and $Z_{H}$ exchange to this
process.

The production cross section $\sigma(S)$ of the process
$e^{+}e^{-}\rightarrow \nu\overline{\nu}W^{*}W^{*}\rightarrow
\nu\overline{\nu}t\overline{t}$ is dominated by collisions of two
longitudinal W's at the parton level. In the following, we will
first discuss the contributions of $B_{H}$ exchange to this
process and see whether the possible signals of $B_{H}$ can be
detected at the future $ILC$ experiments with $\sqrt{S}=1TeV$. So,
as numerical estimation, we will focus our attention on the
production of the top quark pairs via longitudinal gauge boson
$WW$ fusion. The main "non-standard" parts of the cross section
for the subprocess $W^{+}_{L}W^{-}_{L}\rightarrow t\overline{t}$
generated by $B_{H}$ exchange can be written as:
\begin{eqnarray}
\hat{\sigma}_{BB}(\hat{S})&=&
\frac{25\pi\alpha^{2}}{32s_{W}^{4}}\frac{\nu^{4}}{f^{4}}(c'^{2}-s'^{2})^{2}
\{[\frac{5}{6}(\frac{2}{5}-c'^{2})-\frac{1}{5}x_{L}]^{2}(3-\beta_{t}^{2})
 \nonumber\\
&&+2[\frac{1}{5}-\frac{1}{2}c'^{2}-\frac{1}{5}x_{L}]^{2}\beta_{t}^{2}\}\beta_{t}
\cdot\frac{\hat{S}^{3}}{m^{4}_{W}}X_{B}^{2},
\end{eqnarray}
\begin{equation}
\hat{\sigma}_{B\gamma}(\hat{S}) =
\frac{5\pi\alpha^{2}}{4s_{W}^{2}}\frac{\nu^{2}}{f^{2}}
(c'^{2}-s'^{2})[\frac{5}{6}(\frac{2}{5}-c'^{2})-\frac{1}{5}x_{L}]
\beta_{t}(-3+\beta_{t}^{2})\cdot
\frac{\hat{S}^{2}}{m^{4}_{W}}X_{B},
\end{equation}
\begin{eqnarray}
\hat{\sigma}_{BZ}(\hat{S}) &=& \frac{5\pi\alpha^{2}}{32s_{W}^{4}}
\frac{\nu^{2}}{f^{2}}(c'^{2}-s'^{2})
\{-(1-\frac{8}{3}s_{W}^{2})[\frac{5}{6}(\frac{2}{5}-c'^{2})-\frac{1}{5}x_{L}]
(3-\beta_{t}^{2})\nonumber\\
&&+2[\frac{1}{5}-\frac{1}{2}c'^{2}-\frac{1}{5}x_{L}]
\beta_{t}^{2}\}\frac{\hat{S}^{3}}{m^{4}_{W}}X_{B}X_{Z},
\end{eqnarray}
\begin{eqnarray}
\hat{\sigma}_{B b}(\hat{S})&=&\frac{15\pi\alpha^{2}}{16s_{W}^{4}}
\frac{\nu^{2}}{f^{2}}(c'^{2}-s'^{2})
[\frac{5}{6}(\frac{2}{5}-c'^{2})-\frac{1}{5}x_{L}]
[-\frac{4}{3}\beta_{t}^{2}-\frac{1-\beta_{t}^{4}}{2}\nonumber\\
&&+\frac{(1-\beta_{t}^{2})^{3}}{4\beta_{t}}L]
\beta_{t}\frac{\hat{S}^{2}}{m^{4}_{W}}\cdot X_{B}
\end{eqnarray}
with
\begin{eqnarray}
X_{i}=\frac{\hat{S}-M_{i}^{2}}{(\hat{S}-M_{i}^{2})^{2}+M_{i}^{2}\Gamma_{i}^{2}},
\end{eqnarray}
in which $\Gamma_{i}$ is the total decay width of the gauge boson
$Z$ or $B_{H}$. $\hat{\sigma}_{ij}(\hat{S})(i\neq j)$ denotes the
interference cross section of the $i$ and $j$ intermediate states.

\begin{figure}[htb] \vspace*{-0.5cm}
\begin{center}
\epsfig{file=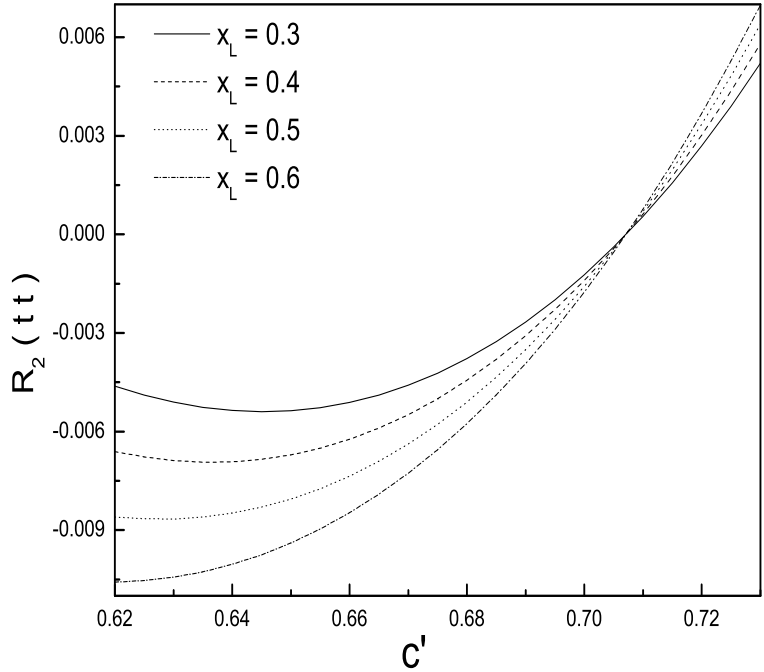,width=320pt,height=300pt} \vspace{-1cm}
\hspace{0.5cm} \caption{The parameter $R_{2}(t\bar{t})$ as a
function of the mixing parameter $c'$ for $f=1TeV$
\hspace*{1.8cm}and four values of the $x_{L}$.} \label{ee}
\end{center}
\end{figure}

We use the relative correction parameter $R_{2}(t\overline{t})=
\Delta\sigma_{2}(t\overline{t})/ \sigma_{2}^{SM}(t\overline{t})$
to represent the contributions of $B_{H}$ exchange to the process
$e^{+}e^{-}\rightarrow
W^{*}_{L}W^{*}_{L}\nu\overline{\nu}\rightarrow
\nu\overline{\nu}t\overline{t}$, in which $\Delta\sigma_{2}
(t\overline{t})$ denotes the corrections of $B_{H}$ exchange to
the $SM$ cross section $\sigma_{2}^{SM}(t\overline{t})$. In Fig.8,
we plot $R_{2}(t\bar{t})$ as a function of the mixing parameter
$c'$ for $f=1TeV$ and four values of the mixing parameter $x_{L}$.
One can see from Fig.8 that the contributions of the heavy gauge
boson $B_{H}$ to the process $e^{+}e^{-}\rightarrow
W^{*}W^{*}\nu\bar{\nu}\rightarrow\nu\bar{\nu}t\bar{t}$ depend
rather significantly on the mixing parameter $c'$. The value of
the relative correction parameter $R_{2}(t\overline{t})$ is
positive or negative, which depends on the value of  the mixing
parameter $c'$. However, its value is very small,
$|R_{2}(t\overline{t})|\leq 1\% $, in all of the parameter space
allowed by the electroweak precision constraints. Thus, the
possible signals of the gauge boson $B_{H}$ can not be studied via
the process $e^{+}e^{-}\rightarrow
W^{*}W^{*}\nu\bar{\nu}\rightarrow\nu\bar{\nu}t\bar{t}$ in the
future $ILC$ experiments.

From Eqs.(24)-(28), we can see that the contributions of the heavy
photon $B_{H}$ to the production cross section for the process
$e^{+}e^{-}\rightarrow
W^{*}W^{*}\nu\bar{\nu}\rightarrow\nu\bar{\nu}t\bar{t}$ are mainly
proportional to the factors $\nu^{4}/f^{4}$ and
$1/(\hat{S}-M_{B}^{2})^2$ from pure $B_{H}$ contributions and to
the factors $\nu^{2}/f^{2}$ and $1/(\hat{S}-M_{B}^{2})$ from the
inference with non-$B_{H}$ contributions. Furthermore, the gauge
boson $B_{H}$ mass $M_{B_{H}}$ is proportional to $f$ for the
fixed value of $c'$. To see the effects of the scale parameter $f$
on the relative correction parameter $R_{2}(t\overline{t})$, we
plot $R_{2}(t\overline{t})$ as a function of $f$ for
 $x_{L}=0.5$ and three values of the mixing parameter $c'$. One can
see from Fig.9 that the deviation of the production cross section
from its $SM$ value is also very small.

\begin{figure}[htb]\vspace{-0.5cm}
\begin{center}
\epsfig{file=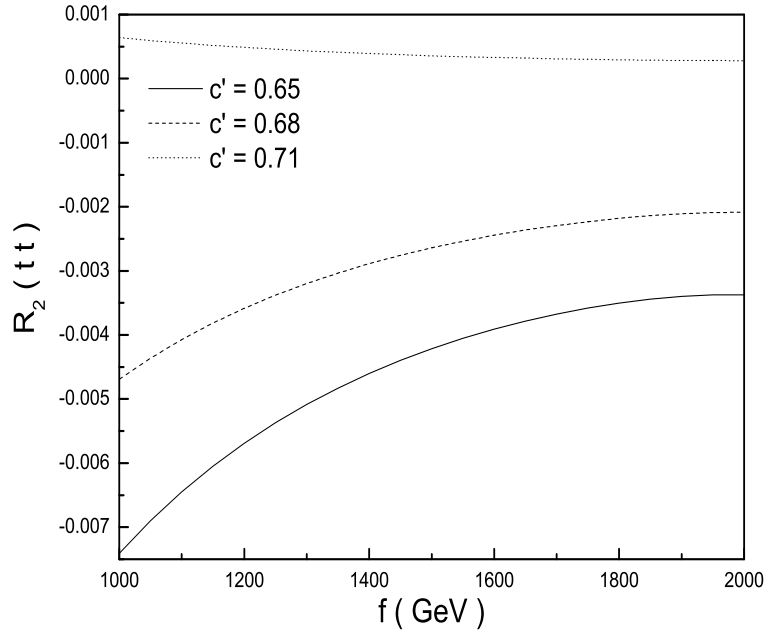,width=320pt,height=280pt} \vspace{-1cm}
\hspace{0.5cm} \caption{The parameter $R_{2}(t\bar{t})$ as a
function of the scale parameter $f$ for $x_{L}=0.5$ and
\hspace*{1.8cm}three values of the parameter $c'$.} \label{ee}
\end{center}
\end{figure}

Similarly to calculation for the contributions of $B_{H}$ exchange
to the process $e^{+}e^{-}\rightarrow W^{*}W^{*}\nu\bar{\nu}
\rightarrow\nu\bar{\nu}t\bar{t}$, we can write the expressions of
the "non-standard" parts of the production cross section for this
process generated by $Z_{H}$  exchange and calculate the relative
deviation of the cross section from its $SM$ value. However, our
numerical results show that the contributions are also very small
in all of the parameter space allowed by the electroweak precision
constraints. Thus, the possible signals of the gauge bosons
$Z_{H}$ and $B_{H}$ can not be detected via the process
$e^{+}e^{-}\rightarrow W^{*}W^{*}\nu\bar{\nu}
\rightarrow\nu\bar{\nu}t\bar{t}$ in the future $ILC$ experiments.

\noindent{\bf VI. Conclusions and discussions}

Little Higgs theory revives an old idea to keep the Higgs boson
naturally light. In all of little Higgs models, there is a global
symmetry structure that is broken at a scale $f$ to make the Higgs
particle as a pseudo-Goldstone boson. In general, these models
predict the existence of the new heavy gauge bosons, new heavy
fermions, and some of heavy triplet scalars. In this paper, we
discuss the possible signals for some of these new particles
predicted by the $LH$ model via studying the $WW$ fusion processes
at the future $ILC$ experiments with $\sqrt{S}=1TeV$.

A future $ILC$ will measure the production cross section of a
light Higgs boson in Higgs-strahlung or $WW$ fusion with
percent-level precision[4]. In particular, $WW$ fusion is the
dominant contribution to Higgs production for $m_{H}< 180GeV$ at
the $ILC$ experiments with $\sqrt{S}\geq 500GeV$. We study the
production of the light Higgs boson from the $WW$ fusion process
$e^{+}e^{-}\rightarrow W^{*}W^{*}\nu\bar{\nu}
\rightarrow\nu\bar{\nu}H^{0}$ in the context of the $LH$ model and
calculate the deviation of the production cross section from its
$SM$ value at the future $ILC$ with $\sqrt{S}=1TeV$. We find that
the value of the relative correction parameter $R(H^{0})$ is
larger than $5\%$ over a sizable region of the parameter space
preferred by the electroweak precision data, which is comparable
to the future $ILC$ measurement precision.

In the $SM$, the process $W^{+}W^{-}\rightarrow t\overline{t}$ can
be generated via the $t-$channel $b$ quark exchange and
$s-$channel $\gamma$, $Z$, $H^{0}$ exchanges. The contributions of
the light Higgs boson predicted by the $LH$ model to this process
contain the pure Higgs contributions and the interference with $b$
quark contributions. Using the EWA method, we calculate the
deviation of the production cross section for the process
$e^{+}e^{-}\rightarrow W^{*}W^{*}\nu\bar{\nu}\rightarrow
H^{0}\nu\bar{\nu}\rightarrow\nu\bar{\nu}t\bar{t}$ from its $SM$
prediction. We find that the relative correction can be
significantly large for reasonable values of the parameters in the
$LH$ model. For example, the value of the relative correction
parameter $R_{1}(t\overline{t})$ is larger than $10\%$ for $f\leq
2TeV$ in most of the parameter space, which is consistent with the
electroweak precision constraints. If we use the computer code
NextCalibur to give the exact cross section, then the value of
$R_{1}(t\bar{t})$ is approximately suppressed by a factor $1/2.5$.
The value of $R_{1}(t\bar{t})$ is larger than $10\%$ for the scale
parameter $f\leq 1.5TeV$. Furthermore, the main backgrounds,
$t\bar{t}$ and $e^{+}e^{-} t\bar{t}$ production, to the signal of
the process $e^{+}e^{-}\rightarrow
W^{*}W^{*}\nu\bar{\nu}\rightarrow \nu\bar{\nu}t\bar{t}$ can be
efficiently reduced by the suitably cuts. Thus, the correction
effects can be seen as new signals of the light Higgs boson and
should be detected via this process at the future $ILC$
experiments with $\sqrt{S}=1TeV$.

The heavy gauge bosons $Z_{H}$ and $B_{H}$ can produce the
corrections to the process $e^{+}e^{-} \rightarrow
W^{*}W^{*}\nu\bar{\nu}\rightarrow \nu\bar{\nu}t\bar{t}$. However,
our numerical results show that the correction effects are very
small in all of the parameter space preferred by the electroweak
precision data. Thus, these new particles can not produce
observable signals via this process at the future $ILC$
experiments.

\vspace{1.0cm} \noindent{\bf Acknowledgments}

This work was supported in part by Program for New Century
Excellent Talents in University(NCET), the National Natural
Science Foundation of China under the grant No.90203005 and
No.10475037, and the Natural Science Foundation of the Liaoning
Scientific Committee(20032101).

\null
\end{document}